\documentclass{article}
\usepackage{RR}

\usepackage{mystyle}
\usepackage{ccs}

\def\pR{\sqsubseteq}
\def\CTS{\msf{CTS}}
\def\cA{\prec}
\def\cH{\#}

\def\st{^{\ast}}
\def\mo{^{{-1}}}
\def\mi{^{{-}}}
\def\mcl#1{\mathcal{#1}}
\def\R{\mathrel{\mc R}} 
\def\full{\mit{full}}
\def\prt{\mit{part}}

\def\Conf{\mc C}

\RRdate{June 2006}

\RRauthor{
  Jean Krivine
}
\authorhead{Krivine}

\RRtitle{Un outil de vérification pour la Programmation Concurrente D\'eclarative}

\RRetitle{A verification algorithm for Declarative Concurrent Programming}
\titlehead{Declarative Concurrent Programming}
\RRresume{ Nous proposons une méthode de vérification pour les
  systèmes distribués basé sur la distinction entre comportement avant
  et arrière d'un système transactionnel. Cette méthode utilise un
  algorithme basé sur les structures d'événements qui, étant donné un
  processus CCS, construit son système de transition causal relatif à
  un ensemble d'actions observables. La vérification du processus CCS
  d'origine, équipé d'un mécanisme de retour arrière sur les
  transitions non observables, revient à vérifier la correction du
  système de transitions causales du processus qui est en général
  beaucoup plus petit. Cette méthode est appelée programmation
  concurrente déclarative (PCD).

  Les performances de la PCD comparées aux performances des techniques
  traditionnelles de bisimulation donnent des résultats encourageants.
  Un banc d'essai utilisant le problème classique du dîner des
  philosophes montre que la PCD est plus efficace que la méthode
  directe, à la fois en terme de temps et d'espace de calcul requis.
  En effet, les outils standard de bisimulation peuvent vérifier des
  systèmes allant au delà de 15 philosophes dans le cas de la PCD,
  alors qu'ils ne peuvent gérer plus de 5 philosophes avec un approche
  de programmation directe. Cet amélioration des performances est
  d'autant plus spectaculaire que la taille du système de
  spécification des philosophes est exponentielle dans le nombre
  d'agents.
}%

\RRabstract{ A verification method for distributed systems based on
  decoupling forward and backward behaviour is proposed. This method
  uses an event structure based algorithm that, given a CCS process,
  constructs its causal compression relative to a choice of observable
  actions.  Verifying the original process equipped with distributed
  backtracking on non-observable actions, is equivalent to verifying
  its relative compression which in general is much smaller. We call
  this method Declarative Concurrent Programming (DCP).

  DCP technique compares well with direct bisimulation based methods.
  Benchmarks for the classic dining philosophers problem show that
  causal compression is rather efficient both time- and space-wise.
  State of the art verification tools can successfully handle more
  than 15 agents, whereas they can handle no more than 5 following the
  traditional direct method; an altogether spectacular improvement,
  since in this example the specification size is exponential in the
  number of agents.  }
\RRkeyword{Process algebra, transaction, event structures,
  verification, bisimulation}
\RRmotcle{Algèbres de processus, transactions, structures d'événements, vérification, bisimulation}
\RRprojet{MOSCOVA}
\URRocq 

\begin{document}
\makeRR   

\section{Introduction}
Backtracking is commonplace in transactional systems where different
components, such as
processes accessing a distributed database, need to acquire a resource simultaneously.  To ensure unconditional
correctness of the overall execution of the transaction, one usually
provides a code that incorporates explicit escapes from those cases
where a global consensus cannot be met. Such an upfront method
generates a large and unstructured state space, which often means
verification based on proving that the code is bisimilar to a
reference specification becomes unfeasible.

Based on earlier work, we propose here an indirect verification
method, and show on an example that it can handle larger
specifications.  The idea is to break down the distributed implementation of a given reference specification in two steps.
First, one writes down a code which is only required to meet a weaker
condition of causal or forward correctness relative to the specification. This condition is parameterised by a choice of observable actions corresponding to the actions of the specification. Second,
the obtained code is equipped with a generic form of distributed
backtracking on non-observable actions. A general theorem reduces the correctness of the latter
partially reversible code to the causal correctness of the
former~\cite{DanKri05}. 


In many transactional examples, this structured programming method works well, and obtains codes which are smaller, and simpler to understand~\cite{DanKriTar05}. It also seems interesting from a correctness perpective, since one never has to deal with the full state space, and it is enough to consider the much smaller state space of
the forward code causal compression relative to observable actions.  
Thus it obtains codes which are also
easier to prove correct. It is only natural then to ask whether and to
which extent such indirect correctness proofs can be automated.
This is the question we address in this paper.



Specifically we propose
an algorithm, which, under certain rather mild assumptions about the
system of interest, will compute its causal compression relative to a
choice of observables. The true concurrency semantics tradition of using event structures as an intrinsic process representation comes to the rescue here. Indeed,
event structures provide a representation of computation traces up to
trace equivalence, and therefore reduce redundancy during the search
of the compression. Besides event structures are uniquely suited to
the handling of causal relationships between various events triggered
by a process~\cite{Win82}. For these reasons our procedure includes a  translation of the process as a recursive flow event structure, and
computes the relative causal compression on this intermediate representation.
%
The algorithm also relies on a compact representation of the conflict relation between events, and seems to perform well both space-wise, obtaining a much smaller state space, and time-wise. Benchmarks given for the classical example of the dining philosophers show a significant state compression, and a relatively low cost incurred by compression. Direct programming
generates a state space that is already too big for being constructed
by bisimulation verifiers for 6 agents, whereas our method can go well
beyond 15.

The language we use to formalize concurrent systems is the
Calculus of Communicating Systems (CCS)~\cite{Mil89}. This is a slightly
more expressive language than basic models of communicating automata, in that
processes can dynamically fork. On the other hand, this communication model includes no name-passing, which is a severe limitation in some applications.
As is discussed further in the conclusion it is possible to adapt the present development, which is largely independent of the chosen communication model, to richer languages such as $\pi$-calculus. 

Section~2 starts with a quick recall of CCS~\cite{Mil89}. Section~3
develops its reversible variant RCCS, together with the central notion of causal correctness, and the fundamental result connecting causal correctness of a CCS process and full correctness of its lifting as a partially reversible process in
RCCS~\cite{DanKri05}. The relative causal compression algorithm, and the accompanying verification method are explained in Section~4. Section~5 compares this method with the traditional direct method,  using the dining philosphers problem as a benchmark. The conclusion discusses related work and further directions.

\section{CCS}\label{sec:CCS}
\subsection{Syntax}
CCS processes interact through binary communications on named
channels: an output on channel $x$ is written $\bar x$, an input on the same channel is simply written $x$. The complete syntax is given in Fig~\ref{ccs}.
\begin{figure}[t]
\centering
\AR{
  \mbox{Processes}&p,q&::=&a.p &\hfill \mbox{Action prefixing}\\
  &&&p\mid q&\hfill\mbox{Parallel composition}\\
  &&&p+q&\hfill\mbox{Choice}\\
  &&&D(\tilde x):=p&\hfill\mbox{Recursive definition}\\
  &&&(x)p&\hfill\mbox{Name restriction}\\
  &&&0&\hfill\mbox{Empty process}\\\\
  \mbox{Actions}&a&::=&x,y,\dots&\hfill\mbox{Input}\\
  &&&\bar x,\bar y,\dots&\hfill\mbox{Output}\\
  &&&\tau&\hfill\mbox{Silent action}
}
   \caption{CCS syntax}
   \label{ccs}
\end{figure}

We write $P$ for the set of processes, $A$ for the set of actions, and
$A\st$ for the free monoid of action words. Restriction $(x)p$ binds
$x$ in $p$ and the set of free names of $p$ is defined accordingly. In
a recursive definition $D(\tilde x):=p$ free names of $p$ have to be
$\tilde x$.


\subsection{Operational semantics}
A \emph{labelled transition system} (LTS) is a tuple
$\tuple{S,s,L,\rar}$ where $S$ is called the state space, $s$ the
initial state, $L$ the set of labels, and $\rar\ \subseteq S\times
L\times S$ the transition relation. One uses the common notation $s\rar_a t$, and for $m=a_1\ldots a_n\in A\st$,
$s\rar\st_m t$ means $s\rar_{a_1}s_1$, \dots, ${s_{n-1}}\rar_{a_n}t$
for some states $s_1$, \dots, $s_{n-1}$.


The operational semantics of a CCS term $p$ is given by means of such
an LTS $(P,p,A,\rar)$, written $\msf{TS}(p)$, where $\rar$ is given
inductively by the rules in Fig~\ref{ccsop}.
\begin{figure}[t]
   \centering
\AR{
  \begin{array}{c}
  \dfrac{}{a.p+q\rar_a p}\msf{(act)}\\\\
  \dfrac{p\rar_a p'\qquad q\rar_{\bar a}q'}{p\mid q\rar_\tau p'\mid q'}\msf{(synch)}\qquad
  \dfrac{p\rar_a p'}{p\mid q\rar_a p'\mid q}\msf{(par)}\\\\
  \dfrac{p\rar_a p'\quad a\not\in\set{x,\bar x}}{(x)p\rar_a (x)p'}\msf{(res)}
  \qquad
  \dfrac{p\equiv p'\rar_a q'\equiv q}{p\rar_a q}\msf{(equiv)}
  \end{array}
} 
   \caption{CCS labelled transition system}
   \label{ccsop}
\end{figure}
The equivalence relation $\equiv$ is the classical structural congruence for choice and parallel composition, together with the recursion unfolding rule
$\inpar{D(\tilde x):=p}\equiv p$.

\subsection{Process equivalence}
Several variants of observational equivalence for CCS processes have
been considered. We use here a variant of {weak bisimulation} based on the choice of a countable distinguished subset $K$ of the set of actions $A$, which we fix here once and for all. Actions in $K$
are called \emph{observable} actions. The complement $A\setminus K$ of non-observable actions is denoted by $K^c$ and also taken to be countable.  

Let $\mcl
S_1=(S_1,s_1,A,\rar)$ and $\mcl S_2=(S_2,s_2,A,\rar)$ be LTSs both with
labels in $A$, a relation $\R$ over $S_1\times S_2$ is said to be a
\emph{weak simulation} between $\mcl S_1$, $\mcl S_2$, if $s_1\R s_2$
and whenever $p_1\R p_2$: \\--- if $p_1\rar_a q_1$, $a\in K^c$, then
$p_2\rar\st_m q_2$ with $m\in(K^c)\st$, and $q_1\R q_2$; \\--- if
$p_1\rar_a q_1$, $a\in K$, then $p_2\rar\st_m q_2$ with $m\in(K^c)\st
a(K^c)\st$, and $q_1\R q_2$.

The idea is that $\mcl S_2$ has to simulate the behaviour of $\mcl S_1$
regarding observable actions, but is free to use any sequence of non observable
ones in so doing.
Such a relation $\mcl R$ is said to be a \emph{weak bisimulation} if both $\mcl R$ and its inverse $\mcl R\mo$ are weak simulations.  When there is such a
relation, $\mcl S_1$ and $\mcl S_2$ are said to be \emph{bisimilar},
and one writes $\mcl S_1\sim\mcl S_2$.  

A CCS process $p$ is said to
be a \emph{correct implementation} of a specification LTS $\mcl S$, if $\msf{TS}(p)\sim\mcl S$. When the specification is clear from the context, we may simply say $p$ is correct. One thing to keep in mind is that
all these definitions are relative to a choice of $K$. Usually, $K$ is
taken to be $A\setminus\set{\tau}$, but this more flexible definition
will prove convenient.

\section{Reversible CCS}\label{sec:RCCS}
We turn now to a quick intuitive introduction to RCCS.  Consider the
following CCS process: \EQ{\label{eq:dead} (x)\inpar{x\mid x\mid\bar
    x.\bar x.a.p\mid\bar x.\bar x.b.q} } Both subprocesses $a.p$ and
$b.q$ require two communications on $x$ to execute, so the whole
process may reach a deadlocked state $(x)\inpar{\bar x.a.p\mid \bar
  x.b.q}$ where neither $a$ nor $b$ may be triggered.
If the intention is that the system implements
the mutual exclusion process $a.p+b.q$, a possible fix is
to give both subprocesses the possibility to release $x$:
\EQ{
\label{eq:full} 
(x)\inpar{x\mid x\mid R_p(x,a)\mid R_q(x,a)}
}
with 
$R_p(x,a):=\bar x.\inpar{ \tau.(R_p(x,a)\mid x)+ \bar
  x.(\tau.(R_p(x,a)\mid x\mid x)+ a.p)} $.

This example helps in realising two key things: first the original
code~(\ref{eq:dead}) although not correct, is partially correct in the
sense that
any successful action $a$ or $b$ leads to a correct state $p$ or $q$;
second the proposed fix can be made an instance of a generic
distributed backtracking mechanism. The idea of RCCS is to provide
such a mechanism, in a way that partial or causal correctness (yet to be
defined formally) in CCS, can be proved to be equivalent to full correctness of
the same process once lifted to RCCS~\cite{DanKri04}.

\subsection{Syntax}
RCCS \emph{forward actions} are the same actions as CCS, namely $A$.
Recall these are split into $K$ and its complement
$K^c$. In the RCCS context actions in $K$ are also called irreversible, or sometimes commit actions (following the transaction terminology);  
actions in $K^c$ are also called reversible, since these are the ones one wants to backtrack. RCCS therefore 
also has \emph{backward actions} written $a\mi$, with $a\in K^c$.

RCCS processes are composed of \emph{threads} of the form $m\trgr p$,
where $m$ is a \emph{memory}, and $p$ is a plain CCS process:
\AR{r::=m\trgr p\mid (r\mid r)\mid (x) r} Memories are stacks used to
record past interactions: \AR{m::=\mem{\th,a,p}\cdot m\mid
  \memc{\th}\cdot m\mid\emem} where $\ta$ is a thread identifier drawn
from a countable set. Open memory elements
$\mem{\th,a,p}$ are used for reversible actions and contain a thread
identifier $\th$, the action last taken, and the alternative process
that was left over by a choice if any. Closed memory elements
$\memc{\th}$ are used for irreversible actions, and only contain an
identifier. The prefix relation on memories is defined as $m\pR m'$ if
there is an $m''$ such that $m''\cdot m=m'$. 

Processes are considered
up to the usual congruence for parallel composition together with the
following specific rules:%
\AR{
  m\trgr\inpar{D(\tilde x):=p}&\equiv& m\trgr p&\\
  m\trgr (p\mid q)&\equiv& (m\trgr p)\mid (m\trgr q)& \\
  m\trgr (x)p &\equiv &(x)(m\trgr p)&\mbox{if } x\not\in m } Any CCS
process $p$ can be lifted to RCCS with an empty memory
$\ell(p):=\emem\trgr p$, and conversely, there is a natural forgetful
map $\phi$ erasing memories and mapping back RCCS to CCS. Clearly
$\phi(\ell(p))=p$. When we want to insist that the lift operation is
parameterised by the set $K$, we write $\ell_K(p)$.

\subsection{Operational semantics}
The operational semantics of RCCS is also given as an LTS with transitions
given inductively by the rules in Fig~\ref{rccsop}.
\begin{figure}[t]
   \centering
\AR{
  \begin{array}{c}
    \dfrac{a\in K^c\quad\th\not\in m}{m\trgr a.p+q\rar_a^\th \mem{\th,a,q}\cdot m\trgr p}\msf{(act)}
    \qquad
    \dfrac{a\in K^c\quad}{\mem{\th,a,q}\cdot m\trgr p\rar_{a}^{\th\mi} m\trgr a.p+q }\msf{(act\st)}\\\\
    \dfrac{a\in K\quad\th\not\in m}{m\trgr k.p+q\rar_k^\th\memc{\th}\cdot m\trgr p}\msf{(commit)}
\\\\ 
\dfrac{r\rar_a^{\Theta}r'\quad\th\not\in s}{r\mid
    s\rar_a^{\Theta}r'\mid s}\msf{(par)}\qquad %
  \dfrac{r\rar_a^{\Theta}r'\quad s\rar_{\bar a}^{\Theta}s'}{r\mid
    s\rar_\tau^{\Theta}r'\mid s'}\msf{(synch)}
  \\\\
  \dfrac{r\rar_a^{\Theta}r'\quad a\neq x,\bar x}
  {(x)r\rar_a^{\Theta}r'}\msf{(res)}\qquad \dfrac{r\equiv
    r'\rar_a^{\Theta}s'\equiv s}{r\rar_a^{\Theta}s}\msf{(equiv)}%
  \end{array}
}%
   \caption{RCCS labelled transition system}
   \label{rccsop}
\end{figure}
In the contextual rules $\Theta$ stands either for $\th$ or $\th\mi$. 
The freshness of the thread identifier $\ta$ is
guaranteed by the side conditions $\ta\not\in m$ in the \textsf{(act)}
and \textsf{(commit)} rules, and $\ta\not\in s$ in the \textsf{(par)}
rule. 
The use of such identifiers makes the presentation given here somewhat simpler than the earlier one~\cite{DanKri05}. Note that backtracking as defined in the operational semantics is a binary communication mechanism of exactly the same nature as usual forward communication. However, since threads are required to backtrack with the exact same thread with which they communicated earlier, 
backtrack can be shown to be confluent, at least for those processes that
are reachable from the lifting of a CCS process.

The \textsf{(commit)} rule uses a closed memory element $\memc{\th}\cdot m$ indicating that the information contained in $m$ is no longer needed, since 
by definition actions in $K$ are not backtrackable.  
Supposing $r$ is a process where any recursive process definition is guarded by a commit, an assumption to which we will return later on, this bounds the total size of open memory elements in any process reachable from $r$.

\subsection{The fundamental property}
%
The question is now to see whether it is possible to obtain a characterisation of the behaviour of a lifted process $\ell_K(p)$ solely in terms of $p$. 
Intuitively, $\ell_K(p)$ being $p$ enriched with a mechanism for escaping computations not leading to any observable actions, one might think
that $\ell_K(p)$ is bisimilar to the transition system generated by those traces
of $p$ which lead to an observable action. This is almost true.


To give a precise statement, we need first a few notations and definitions. An RCCS transition as
defined above is fully described by a tuple $t=\tuple{r,a,\Theta,r'}$
where $r$ is the \emph{source} of $t$, $r'$ its \emph{target}, $a$ its
label and $\Theta$ its identifier. If $a\in K$ we say that $t$ is a
\emph{commit} transition, otherwise it is a \emph{reversible}
transition.  If $\Th=\th$ ($\Th=\th\mi$) we say $t$ is \emph{forward}
(\emph{backward}).  A \emph{trace} is a sequence of composable
transitions, and we write $r\rar\st_\sig s$ ($p\rar\st_\sig q$)
whenever $\sig$ is an RCCS (CCS) trace with source $r$ ($p$) and
target $s$ ($q$). A trace is said to be forward if it contains only
forward transitions.

A final and key ingredient is the notion of causality between
transitions in a given forward trace. For CCS this is usually defined
using the so-called proof terms~\cite{BouCas89}, but one can also use
RCCS memories.

The set of memories involved in a forward transition
$t=\tuple{r,a,\th,r'}$ is defined as $\mu(t):=\set{m\in r\mid \exists
  a,q:\mem{\th,a,q}.m\in r'}$; this is either a singleton, if no
communication happened, or a two elements set, if some did.%
\DEF{
  [Causality] Let $\sig:t_1;\dots;t_n$ be a forward RCCS trace:
  \\--- $t_i$ and $t_j$ with $i<j$, are in \emph{direct
    causality relation}, written $t_i<_1t_j$ if there is
  $m\in\mu(t_i)$, $m'\in\mu(t_j)$ such that $m\sqsubset m'$;
  one says that $t_i$ \emph{causes} $t_j$, written $t_i<t_j$, if
  $t_i<_1^* t_j$.
  \\--- $\sig$ is said to be \emph{causal} if for all transitions $t_i$
  with $i<n$, $t_i<t_n$; it is said to be \emph{$k$-\-causal} 
  if it is causal, its last transition $t_n$ is labelled with $k\in K$, 
  and all preceding transitions are labelled in $K^c$.
  }%
One extends this terminology to CCS traces by saying a CCS trace $p\rar^*_\sig p'$ is causal, if it lifts to a \emph{causal} trace $\ell_K(p)\rar^*_{\sig'} r'$ with $\phi(r')=p'$.




With the notion of causal trace in place, we can define the causal
compression of a process $p$ relative to $K$.%
\begin{definition}[Relative causal compression] Let $p$ be a CCS process, its
  \emph{causal compression relative to $K$}, written $\CTS_K(p)$, is the LTS
  $\tuple{P,p,K,\crar}$ where $\crar_k$ is
  defined as $q\crar_k q'$ if $q\rar^*_\sig q'$ for some $k$-\-causal
  trace $\sig$.
\end{definition}
We are now ready to state the theorem that characterizes the behaviour
of $\ell_K(p)$ in terms of the simpler process $p$.
\begin{theorem}[Fundamental property~\cite{DanKri05}]
\label{thm:cts} Let $\msf{TS}_K(p):=\tuple{R,\ell_K(p),A,\rar}$ be
  the LTS associated to the lift $\ell_K(p)$, $\msf{TS}_K(p)\sim\CTS_K(p)$.
\end{theorem}
As said above, it is not true that $\msf{TS}_K(p)$ is bisimilar
to the transition system of traces of $p$ leading to observable actions,
one has to be careful to restrict to causal traces. A trivial but useful rephrasing of this result is:
\begin{corollary}\label{coro}
  Let $p$ be a CCS process, and $\mc S$ be its specification,
  if $\CTS_K(p)\sim\mc S$ then $\ell_K(p)\sim\mc S$.
\end{corollary}
In words, this says that to check the correctness of $\ell_K(p)$ with
respect to $\mc S$, it is enough to check the correctness of
$\CTS_K(p)$. 

If one goes back to the example at the beginning of this section,
this says that $\ell_{\set{a,b}}((x)\inpar{x\mid x\mid \bar x.\bar
x.a.p\mid \bar x.\bar x.b.q})$ is equivalent to
$a.p+b.q$, as soon as the causal compression
of $p=(x)\inpar{x\mid x\mid \bar x.\bar x.a.p\mid \bar x.\bar x.b.q}$ 
relative to $\set{a,b}$ is. This is easily seen in this example, and in fact, as often in practice, $\CTS_K(p)$ and $\mc S$ turn out to be equal.

The interest of this fundamental property lies in the fact that the causal compression relative to $K$, $\CTS_K(p)$, is significantly smaller than the partially reversible process $\ell_K(p)$. A natural question is therefore, given a process $p$, to compute $\CTS_K(p)$. By finding an efficient way to do this, one would obtain an efficient verification procedure. 
This is the object of the next section.

\section{Causal compression}
A first idea to extract the causal transition system of a process $p$
is to use the LTS generated by $\ell(p)$ and screen off non causal traces. 
One cannot know however whether a trace can be extended into a $k$-causal form until a commit is effectively taken, and such an approach would likely 
lead to both superfluous (because lots of traces will
not be causal) and redundant (because of trace equivalence)
computations. A more astute approach is to look only at traces that
will eventually be in a $k$-causal form.  This requires a bottom up
view of traces where one starts from commits inside a term, 
and then reconstructs causal traces triggering
this commit by consuming its predecessors in every possible way. 

However, there is no need to work directly in the syntax, and event
structures~\cite{Win82} provide exactly what is needed here: a truly concurrent semantics that
abstracts from the interleaving of concurrent transitions, and more importantly an explicit notion of causality. Among the
various types of {event structures} the most often considered are
prime ones, because consistent runs can be simply characterized. Yet
they lead to quite large data structures.\footnote{Specifically in
  prime event structure causes of an event must be uniquely
  determined, and this forces duplication of the future of an event
  each time it is engaged in a synchronization.}  Our algorithm uses
instead \emph{flow event structures
  (FES)}~\cite{BouCas89,Bou90,GlaGol03}. On the one hand, there is a
simple inductive translation of CCS terms into FESs that incurs no computational
cost; on the other hand, FES are algorithmically convenient compact
forms of event structures.

We first explain how to extract the causal compression $\msf{CTS}_K(p)$ 
from the translation of $p$ into an FES. Then we
discuss computational issues such as how to make this an algorithm,
and how some of the apparent computational costs can be circumvented
at the level of the implementation.
\subsection{Flow event structures}
A (labelled) flow event structure is a tuple $\mc E = \tuple{E,\cA,\cH,\la}$ where
\\--- $E$ is a set of \emph{events},
\\--- $\cA\ \subseteq E\times E$ is the \emph{flow relation} which has to be irreflexive,
\\--- $\cH\ \subseteq E\times E$ is the \emph{conflict relation} which is symmetric,
\\--- and $\la:E\to A$ a labelling function.
\\
The idea is that the flow relation gives all immediate possible causes of an
event, while the conflict relation indicates a conflicting choice
between two events. 
\DEF{ 
Let $\mc E=\tuple{E,\cA,\cH,\la}$ be an FES, a set
$X\subseteq E$ is a \emph{configuration} of $\mc E$, 
written $X\in\Conf(\mc E)$, if it is:
  \\--- \emph{conflict free}: $\cH\cap(X\times X)=\nil$,
  \\--- \emph{cycle free}: $\cA\st/X$ is a partial order,
  \\--- and \emph{left-closed up to conflicts}: if $e\in X$
    and there is $d\in E$ such that $d\cA e$ then either $d\in X$ or
    there exists $f\in X$ such that $f\cA e$ and $f\cH d$.  
}%
The last two conditions are the price to pay for working with FESs, and are not needed for prime ones. The first one will require some optimised structuring of
the conflict relation, we'll return to this point soon.

\begin{figure}[t]
   \centering
   \input{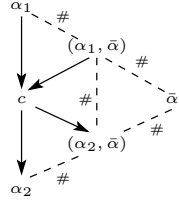} 
   \caption{FES representation of $p:=\al.c.\al.0\mid\bar\al.0$.
     Events are named after their labels when these are not ambiguous.}
   \label{fig:FES}
\end{figure}

A configuration $X$ in $\mc E$ with $e\in X$ is \emph{$e$-minimal} if
$\forall e'\in X:e'\cA^* e$. The set of $e$-minimal configurations is
denoted by $\Conf\tuple{\mc E,e}$.
\\
There is an easy inductive translation $u$ unfolding any CCS process
into a FES~\cite{BouCas89}, where events correspond to communications,
and configurations are those subsets of events that a trace can
trigger. We will not recall here this translation, and only give an
example (see Fig.~\ref{fig:FES}). The correctness of $u$ is given by
the following representation theorem: \THE{[\cite{Bou90}] Let $p$ be a
  CCS process, and $\mc T_\simeq(p)$ stand for the traces of $p$
  quotiented by trace equivalence, then $(\mc T_\simeq(p),\leq)$ and
  $(\Conf(u(p)),\subseteq)$ are isomorphic.  }
One can define a transition system out of an FES. To do this, we define
$\mc E|X$, the \emph{residual} of $\mc E$ by a configuration $X$ in $\Conf(E)$.
\DEF{[Residual] Let $\mc E=\tuple{E,\cA,\cH,\la}$ be an FES, $X$ be a
  configuration of $\mc E$, and define $X_\cH:=\set{e\in E\mid \exists
    e'\in X:e'\cH e}$. The residual of $E$ by $X$ is $\mc
  E|X:=\tuple{E',\cA',\cH'}$ where:%
  \AR{E':=E\setminus(X\cup X_\cH)\quad \cA':=\cA\cap\ (E'\times
    E')\quad \cH':=\cH\cap (E'\times E')} }
The LTS associated to $\mc E=\tuple{E,\cA,\cH,\la}$ has initial state
$\mc E$, and transition relation given by $\mc E'\rar_X \mc E''$ if
$X\in\Conf(\mc E')$ and $\mc E''=\mc E'|X$.  

It is here that our reframing of the compression question in terms of event structures pays off, since to obtain the causal compression of the transition system above, all one has to do is to restrict labels to $e$-minimal configurations such that $\la(e)\in K$. 
The \emph{causal LTS} associated to $\mc E$, written $\CTS_K(\mc E)$,
has initial state $\mc E$, and transition relation given by $\mc
E'\crar_k \mc E''$ if there is an event $e\in E'$ such that $\mc E'\rar_X
\mc E''$ with $X\in\Conf\tuple{\mc E',e}$ and $\la(e)\in K$.
As a consequence of the representation theorem one gets:
\begin{lemma}
Let $p$ be a CCS process, then $\CTS_K(p)$ and $\CTS_K(u(p))$ are isomorphic.
\end{lemma}
At that point, we have an equivalent definition of $\CTS_K(p)$ in terms
of the FES $u(p)$, and it remains to see how one can turn
this definition into an algorithm. This is what we discuss now.

\subsection{Algorithmic discussion}
First, the unfolding $u(p)$ is in general an \emph{infinite} object
even if we restrict to finite state processes. To keep with finite
internal data structures, we require each recursive process definition
to be guarded by a commit action. This seems a reasonable constraint,
in that there is a priori no reason to model a transactional mechanism
with a process that allows infinite forward inconclusive traces.

To compute $\CTS_K(u(p))$, we use instead of $u$, a \emph{partial
unfolding} $u^{\mit{fin}}$ that coincides with $u$ except it does
not unfold any recursive definition. The constraint above ensures that
every commit $k$ that is reachable by a single causal transition can
be seen by this partial unfolding. Only after triggering the event
corresponding to $k$, are the recursive calls guarded by $k$ (if any)
unfolded, and their translations by $u^{\mit{fin}}$ added to the
residual of the obtained event structure. One then checks whether the
obtained residual event structure is isomorphic with some
obtained previously, and adds it to the state space if not.  Given a
process $p$, the algorithm to compute $\CTS_K(u(p))$ proceeds as follows:
\ENU{
\item[0.] $\mc E=\tuple{E,\cA,\cH,\la}:=u^{\mit{fin}}(p)$
\item\label{begin} For all $e\in E$ such that $\la(e)\in K$, compute
  the $e$-minimal configurations $X_e\in \Conf\tuple{\mc E,e}$.
\item For each such $X_e$ build the residual $\mc E|X_e$, with 
  recursive definitions guarded by $e$ unfolded using $u^{\mit{fin}}$.
\item Add the transitions $\mc E\crar_k\mc E|X_e$ to
  the CTS under construction.
\item For each residual $\mc E|X_e$ not isomorphic to any
  previous one, set $\mc E:=\mc E|X_e$ and goto step~\ref{begin}. 
}
By the representation theorem, this algorithm will terminate
as soon as $\CTS_K(p)$ is finite. 

\begin{figure}[ht]
    \begin{center}
      \input{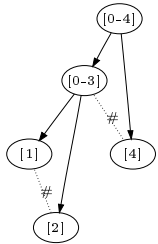}
      \caption{Conflict tree of $a_3.(b_0\mid c_2+d_1) + e_4$}
      \label{fig:confTree}
    \end{center}
  \end{figure}

In practice most of the isomorphism tests can be avoided by using a
quite discriminative equality test between FES signatures which is linear
in the number of events. Another efficiency problem one has 
to deal with is the internal
representation of the conflict relation (which is involved in step 1 because of the conflict-free condition on configurations). In prime event structures
conflict is inherited by causality, that is to say if $e\cH e'$ and $e'\cA e''$, then $e\cH e''$. Hence a rather compact way to represent conflict is to keep
only $(e,e')\in\cH$ and deduce when needed that $e\cH e''$ by
heredity. We have found that a similar compact structure, which we call a 
conflict tree can be used for FESs. Conflict trees are built during process partial unfoldings, and result in a typically logarithmically compact representation of conflict, for a low computational cost. An example of a conflict tree is given Fig.~\ref{fig:confTree}: conflicts is predicated of intervals, and $[n-m]\#[n'-m']$ means that any pair of events indexed within 
$\set{n,\ldots,m}\times\set{n',\ldots,m'}$ is in conflict.

\section{{\sf Causal} module and tests}
The relative compression algorithm was implemented as an
Ocaml~\cite{Ocaml} library {\sf Causal}~\cite{Causal}.
Having a library instead of an independent tool allows to use the underlying
language that offers more construction primitives than CCS.  
Any interesting encoding needs parametric process definitions in order
to define systems with varying number of agents, 
and our module offers simple CCS
process constructors, so that one has a real programming language 
to build large processes.

\subsection{Benchmark}
To get a sense of how well our verification technique performs
compared with a straight bisimulation based verification, we ran
several tests\footnote{Tests were made with an Intel Pentium 4 CPU
  3.20GHz with 1GB of RAM.}  using encodings of the dining
philosophers problem. This timeless example of distributed consensus
involves $n$ philosophers eating together around a table. Each of them
needs two chopsticks to start eating, and has to share them with his 
neighbours. 
When a philosopher
has eaten, he releases his chopsticks after a while and goes
back to the initial state. In the partial implementation, say
$p_\prt$, once a philosopher takes a chopstick he never puts it back
unless he has successfully eaten. In the fully correct one, say
$p_\full$, he may release chopsticks at any time (thus avoiding
deadlocks). The CCS processes $p_\prt$ and $p_\full$ for $n=2$
correspond roughly to the earlier examples (\ref{eq:dead}) 
and (\ref{eq:full}). (See~\cite{DanKri05} for a general definition 
and detailed study.) 

\begin{figure}[ht]
    \begin{center}
      \input{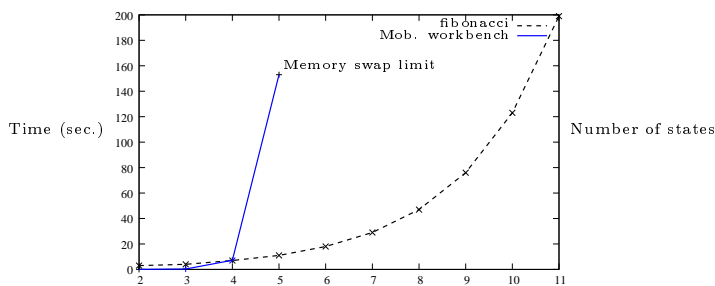}
      \caption{Direct bisimulation test for $p_\full$.}
      \label{fig:bench_mwb}
    \end{center}
  \end{figure}

There are two main reasons for taking the dining philosophers example.
First it is a paradigmatic example of distributed consensus, so the
way to solve it without access to the scheduler (by adding additional
semaphores for instance) has to involve backtracking.  Second, it
turns out that the number of possible states of the specification is
given by a Fibonacci sequence\footnote{Thanks to Hubert Krivine
  (LPTMS) for showing us this nice result.} %
\AR{S(1)=1\quad S(2)=3\quad S(n+1)=S(n)+S(n-1)} This is convenient in
that it gives a simple means to compare the time of computation with
the size of the specification state space. Verifying correctness of
$p_\full$ using the Mobility Workbench (MWB)~\cite{mwb} (see
Fig.~\ref{fig:bench_mwb}) 

\begin{figure}[ht]
    \begin{center}
      \input{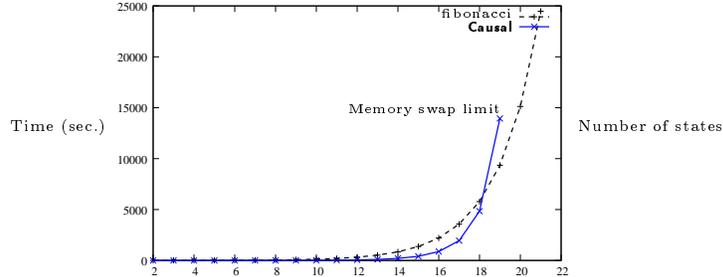}
      \caption{Relative causal compression
using the {\sf Causal} module.}
      \label{fig:bench_causal}
    \end{center}
  \end{figure}

proved to be impossible beyond $5$ philosophers (around $160$
specification states) because of memory limitations.  By using first
the {\sf Causal} module (see Fig.~\ref{fig:bench_causal}) to extract the
causal transition system of $p_\prt$, 

we could verify up to $19$ philosophers (around $15,000$ specification
states) within a time which stayed roughly proportional to the number
of states. Since $\CTS(p_\prt)$ is in this case equal to the specification,
the remaining part of the correctness proof takes negligible
time (MWB needs $0.4s$ for $10$ philosophers).

\section{Conclusion}
We have proposed a method for the verification of distributed systems
which uses an algorithm of relative causal compression. The method
does not always apply: the process one wants to verify must use a
generic backtracking mechanism. This may seem a limitation, but it
often obtains a much simpler code, and many examples of distributed
transactions lend themselves naturally to this constraint. When the
method does apply, however, it proves very effective as we have shown
in the dining philosophers example.

State space explosion in automated bisimulation proofs is a well known
phenomenon, and trace compression techniques have been proposed to
avoid the redundancy created by the interleaving of
transitions~\cite{BouCas89,GodWol91}, and used in model-\-checking
applications~\cite{BiaColDegPri95,Zing}. 
These compressions preserve bisimilarity, whereas our does not, and is
of a completely different nature.  Besides, and because our algorithm
uses event structures, we also cash in on this classical kind of
compression.




There is no reason why this verification method should be limited to
CCS.  Other concurrent models can be equipped with backtracking, and
forward and backward aspects of correctness can be split there as
well.  Recent work extends the concept of partially reversible
computations to various process algebras~\cite{PhiUli06}, and it is
possible to define an analogue of RCCS for the $\pi$-calculus. New
advances in event structure semantics for
$\pi$-calculus~\cite{VarYos06} might allow to extend the causal
compression algorithm, so as to cover the important case of
name-passing calculi.

\small
\bibliographystyle{unsrt}
\bibliography{./concur}

\begin{thebibliography}{10}

\bibitem{DanKri05}
Vincent Danos and Jean Krivine.
\newblock Transactions in {RCCS}.
\newblock In {\em Proceedings of {CONCUR'05}: 16$^{th}$ International
  Conference on Concurrency Theory}, volume 3653 of {\em LNCS}, 2005.

\bibitem{DanKriTar05}
Vincent Danos, Jean Krivine, and Fabien Tarissan.
\newblock Self assembling trees.
\newblock In {\em Proceedings of the $7^{th}$ International conference on
  Artificial Evolution (EA'05)}, 2005.
\newblock To appear.

\bibitem{Win82}
Glynn Winskel.
\newblock Event structure semantics for {CCS} and related languages.
\newblock In {\em Proceedings of 9th ICALP}, volume 140, pages 561--576, 1982.

\bibitem{Mil89}
Robin Milner.
\newblock {\em Communication and Concurrency}.
\newblock International Series on Computer Science. Prentice Hall, 1989.

\bibitem{DanKri04}
Vincent Danos and Jean Krivine.
\newblock Reversible communicating systems.
\newblock In {\em Proceedings of {CONCUR'04}: 15$^{th}$ International
  Conference on Concurrency Theory}, volume 3170 of {\em LNCS}, pages 292--307,
  2004.

\bibitem{BouCas89}
G{\'e}rard Boudol and Ilaria Castellani.
\newblock Permutation of transitions: An event structure semantics for {CCS}
  and {SCCS}.
\newblock In {\em Linear Time, Branching Time and Partial Order in Logics and
  Models for Concurrency}, volume 354 of {\em LNCS}, pages 411--427, 1989.

\bibitem{Bou90}
G{\'e}rard Boudol.
\newblock Flow event structures and flow nets.
\newblock In {\em Proceedings of LITP Spring school on Semantics of Systems of
  Concurrent Processes}, volume 469 of {\em LNCS}, pages 62--95, 1990.

\bibitem{GlaGol03}
Rob van Glabeek and Ursula Goltz.
\newblock Well-behaved flow event structures for parallel composition and
  action refinement.
\newblock {\em Theoretical Computer Science}, 311(1-3):463--478, 2003.

\bibitem{Ocaml}
The {Ocaml} programming language.
\newblock Available at http://caml.inria.fr.

\bibitem{Causal}
Causal --- ocaml module for causality analysis of {CCS} processes.
\newblock Available at http://pauillac.inria.fr/{$\scriptstyle{\sim}$}krivine.

\bibitem{mwb}
Bj{\"o}rn Victor and Faron Moller.
\newblock The {M}obility {W}orkbench --- a tool for the $\pi$-calculus.
\newblock In {\em Proceedings of CAV'94: Computer-Aided Verification}, volume
  818 of {\em LNCS}, pages 428--440, 1994.

\bibitem{GodWol91}
Patrice Godefroid and Pierre Wolper.
\newblock Using partial orders for the efficient verification of deadlock
  freedom and safety properties.
\newblock In {\em Proceedings of {CAV}'91: Computer-aided verification}, volume
  575 of {\em LNCS}, pages 332--342, 1991.

\bibitem{BiaColDegPri95}
Alessandro Bianchi, Stefano Coluccini, Pierpaolo Degano, and Corrado Priami.
\newblock An efficient verifier of truly concurrent properties.
\newblock In {\em Proceedings of Parallel Computing Technologies}, volume 964
  of {\em LNCS}, pages 36--50, 1995.

\bibitem{Zing}
Tony Andrews, Shaz Qadeer, Sriram~K. Rajamani, Jakob Rehof, and Yichen Xie.
\newblock Zing: A model checker for concurrent software.
\newblock In {\em Proceedings of {CAV}'04: Computer-aided verification}, volume
  3114 of {\em LNCS}, pages 484--487, 2004.

\bibitem{PhiUli06}
Iain Phillips and Irek Ulidowski.
\newblock Reversing algebraic process calculi.
\newblock In {\em Proceedings of {FOSSAC}'06}, LNCS, 2006.
\newblock To appear.

\bibitem{VarYos06}
Daniele Varacca and Nobuko Yoshida.
\newblock Typed event structures and the $\pi$-calculus.
\newblock In {\em Proceedings of {MFPS XXII}}, 2006.
\newblock To appear.

\end{thebibliography}

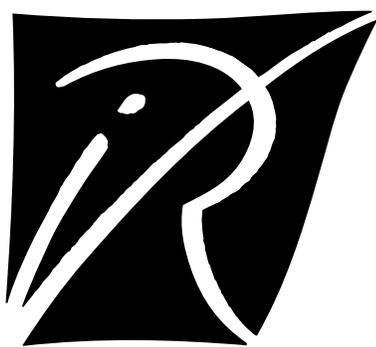
\end{document}